\documentclass[preprint,superscriptaddress,showpacs,preprintnumbers,amsmath,amssymb,aps,pre,floatfix,]{revtex4-1}

\usepackage{graphicx}
\usepackage{dcolumn}
\usepackage{bm}
\usepackage{cancel}
\usepackage{color}
\usepackage[utf8]{inputenc}

\begin{document}

\title{From creation and annihilation operators to statistics}

\author{M. Hoyuelos}
\affiliation{Instituto de Investigaciones F\'isicas de Mar del Plata (IFIMAR -- CONICET), Funes 3350, 7600 Mar del Plata, Argentina}
\affiliation{Departamento de F\'isica, Facultad de Ciencias Exactas y Naturales, Universidad Nacional de Mar del Plata, Funes 3350, 7600 Mar del Plata, Argentina}

\date{\today}

\begin{abstract}
A procedure to derive the partition function of non-interacting particles with exotic or intermediate statistics is presented. The partition function is directly related to the associated creation and annihilation operators that obey some specific commutation or anti-commutation relations. The cases of Gentile statistics, quons, Polychronakos statistics, and ewkons are considered. Ewkons statistics was recently derived from the assumption of free diffusion in energy space (Phys. Rev. E \textbf{94}, 062115, 2016); an ideal gas of ewkons has negative pressure, a feature that makes them suitable for the description of dark energy.
\end{abstract}


\maketitle

\section{Introduction}

Relations between creation and annihilation operators determine the statistical properties of quantum systems composed by non-interacting particles. Canonical examples are the Bose-Einstein distribution for the commutation relation $[a,a^\dagger]=1$ and the Fermi-Dirac distribution for the anti-commutation relation $\{a,a^\dagger\}=1$. Starting from the pioneer works of Gentile \cite{gentile} and Green \cite{green}, many different distributions have been proposed as extensions that go beyond or interpolate the statistics of bosons and fermions; see, for example, \cite{katsura,wilczek,greenberg,mohapatra,haldane,isakov,isakov1,wu,isakov2,poly,kania0,bytsko,anghel,mirza,dai,algin,ardenghi}. Although, according to the Standard Model, fermions and bosons are enough to describe nature from fist principles, there are situations in which a description in terms of exotic statistics is more useful; see \cite[ch.\ 1]{khare} for several examples, and references cited therein including experimental results for some cases.

The total energy is the simple sum $E = \sum_i n_i \epsilon_i$, where $n_i$ is the number of identical particles in energy level $i$, and $\epsilon_i$ is the corresponding one-particle energy. The starting point is the grand partition function written in terms of the number of different many-body states or statistical weight $W(\{n_i\})$:
\begin{equation}
\mathcal{Z} = \sum_{\{n_i\}} W(\{n_i\}) \exp\left[-\beta\sum_i (\epsilon_i - \mu) n_i\right],
\label{zeta}
\end{equation}  
where $\beta=1/k_B T$ and $\mu$ is the chemical potential. A frequent approach to obtain the statistical distribution of $n_i$ for exotic statistics is maximizing $\ln W$ with the constraints of constant total energy and total number of particles, and taking the thermodynamic limit to make some approximations \cite{wu,bytsko}.

Since each energy level can be taken as an independent system, the total grand partition function is written in terms of the single level grand partition functions $\mathcal{Z}_\epsilon$ as
\begin{equation}
\mathcal{Z} = \prod_{\{\epsilon \}} \mathcal{Z}_\epsilon
\end{equation}
(subindex $i$ in $\epsilon$ is omitted to simplify the notation); the product is performed on all energy levels, taking into account a possible degeneracy by repetition of the product. Here, I focus on the grand partition function for a single level given by
\begin{equation}
\mathcal{Z}_\epsilon = \mathrm{tr}\, e^{-\beta\,(\epsilon-\mu) \hat{n}}
\label{zeps}
\end{equation}
where $\hat{n}$ is the number of particles operator. An exotic statistics originated by some specific relations between creation and annihilation operators should manifest itself both in \eqref{zeta} and in \eqref{zeps}. The questions that I wish to address are: how does \eqref{zeps} depend on the creation and annihilation operator relations, and how this dependence, when explicitly stated, could be extended to include exotic statistics without making approximations or appealing to the thermodynamic limit.

As usual, te most convenient base to evaluate the trace in \eqref{zeps} is the set of eigenstates of the number operator:
\begin{equation}
\mathcal{Z}_\epsilon = \sum_n e^{-\beta\,(\epsilon-\mu) n}
\label{zeps2}
\end{equation}

In Sect.\ \ref{gent}, a counting operator is introduced, with eigenstates $|n\rangle$ and eigenvalues 0 or 1, in order to restrict the sum in \eqref{zeps2} to the values of $n$ allowed by some commutation relations. It can be seen that the only possible extension of Eq.\ \eqref{zeps2} beyond fermions and bosons is Gentile statistics; this is one of the main results of Ref.\ \cite{dai}. In order to get other statistics, for example quantum Boltzmann statistics for $a a^\dagger = 1$ \cite{greenberg,mohapatra}, the eigenvalues of the counting operator should be different from 0 and 1. In the next sections this situation is analyzed for several exotic statistics: quons in Sect.\ \ref{squons}, Polychronakos statistics in Sect.\ \ref{spoly} and ewkons in Sect.\ \ref{sewkons}. I present the conclusions in Sect.\ \ref{conclusions}.

\section{Gentile statistics and the counting operator}
\label{gent}

The creation and annihilation operators determine the number of elements of the Fock space and restrict the sum in Eq.\ \eqref{zeps2} to the allowed values of $n$. Then, if $n$ takes the values 0 or 1, we have fermions, if it takes any value between 0 and $\infty$, we have bosons, and if it takes values between 0 and $\nu$ we have an intermediate Gentile statistics. We can define
\begin{equation}
x =  e^{-\beta\,(\epsilon-\mu)}
\end{equation}
in order to simplify the notation. The number distribution is given by
\begin{equation}
\bar{n} = x \frac{\partial\ln \mathcal{Z}_\epsilon}{\partial x} =
\frac{1}{x^{-1}-1} - \frac{\nu + 1}{x^{-(\nu+1)}-1}.
\end{equation}
It reduces to the Fermi-Dirac distribution for $\nu=1$ and to the Bose-Einstein distribution for $\nu \rightarrow \infty$. In this and the next sections I consider creation and annihilation operators whose action on number states is written as
\begin{align}
a^\dagger |n\rangle &= \lambda_{n+1}^* |n+1\rangle \nonumber \\ 
a |n\rangle &= \lambda_n |n-1\rangle,
\label{aadaga}
\end{align}
with the vacuum condition $a|0\rangle = 0$; therefore $\lambda_0 = 0$. For Gentile statistics we have that (see, e.g., \cite{katsura})
\begin{align}
\lambda_n &= \sqrt{n} \quad \text{for } 1 \leq n \leq \nu \nonumber \\
\lambda_{\nu+1} &=0.
\end{align}
The anticommutation relation for fermions that gives $a^2=0$ is generalized to 
\begin{equation}
a^{\nu+1} = 0.
\label{antic}
\end{equation}
If $F_\nu$ is the Fock space represented by the set $\{|0\rangle,...,|\nu\rangle\}$, then any $F_\nu$ is embedded into $F_{\nu'}$ as long as $\nu' > \nu$; and the Fock space of bosons includes all the others.

Let us consider a counting operator $\hat{\delta}$ that commutes with $\hat{n}$ and that has the following property: 
\begin{equation}
\langle n| \hat{\delta} |n\rangle = \left\{ \begin{array}{ll}
1 & \text{if } 0\leq n \leq \nu \\ 
0 & \text{if } n > \nu
\end{array}   \right.
\label{condit}
\end{equation}
The definition of the grand partition function \eqref{zeps} remains unchanged if we insert this operator so that  $\mathcal{Z}_\epsilon = \mathrm{tr}\,(\hat{\delta}\, x^{\hat{n}})$. Now, the sum in \eqref{zeps2} can be extended to infinity:
\begin{equation}
\mathcal{Z}_\epsilon = \sum_{n=0}^\infty \langle n| \hat{\delta} |n\rangle \, x^{n}.
\label{zeps3}
\end{equation}
We now seek to express $\hat{\delta}$ in terms of creation and annihilation operators. For this purpose it is useful to consider relation \eqref{antic}, that gives $a^n |n\rangle =0$ for $n>\nu$. Including ${a^\dagger}^n$ to keep $|n\rangle$ as an eigenstate, and the normalization factor $1/n!$, we obtain that
\begin{equation}
\hat{\delta} = \sum_{n=0}^{\infty} \frac{1}{n!}\, {a^\dagger}^n a^n \, |n\rangle\langle n|
\label{delta}
\end{equation}
satisfies the conditions \eqref{condit}.

I briefly mention a different possible representation and interpretation of $\hat{\delta}$. Let us call $b^\dagger$ and $b$ the creation and annihilation operators for bosons. Then,
\begin{align}
\hat{\delta} &= \sum_{n=0}^{\infty} b^{-n} {b^\dagger}^{-n} {a^\dagger}^n a^n \, |n\rangle\langle n| \nonumber \\
&= \sum_{n=0}^{\infty} \mathcal{N} (b^{-1} {b^\dagger}^{-1} {a^\dagger} a)^n \, |n\rangle\langle n| \nonumber \\
&= \mathcal{N} (b^{-1} {b^\dagger}^{-1} {a^\dagger} a)^{\hat{n}}
\end{align}
where $\mathcal{N}$ is the normal ordering operator; creation and annihilation operators should not be taken as written but in normal order. In the last step it was used that $\hat{n}$ commutes with $a^\dagger a$ and $b^{-1} {b^\dagger}^{-1}$ (in this case, the normal ordering operator does not act on $\hat{n}$). I call $\hat{\gamma} = b^{-1} {b^\dagger}^{-1} {a^\dagger} a$ so that the counting operator is
\begin{equation}
\hat{\delta} = \mathcal{N} e^{(\ln \hat{\gamma}) \hat{n}},
\end{equation}
and including this operator in the grand partition function we have:
\begin{equation}
\mathcal{Z}_\epsilon = \mathrm{tr}\,\left[ \mathcal{N} e^{-\beta\,(\epsilon-\mu - k_B T \ln \hat{\gamma}) \hat{n}} \right].
\label{zeps4}
\end{equation}
Now, we can interpret $\hat{\gamma}$ as a quantum activity coefficient that takes into account quantum effects represented by the features of creation and annihilation operators (the description in terms of quantum operators may also be a way of introducing interaction effects \cite{anghel}). Defined this way, this quantum activity coefficient has a reference state given by a system in which particles are bosons. For simplicity, in the rest of this paper I use the definition of $\hat{\delta}$ given by \eqref{delta} and do not use the operator $\hat{\gamma}$.

So far, the inclusion of the counting operator has no consequence in Gentile statistics. As stated before, it is not possible for the definition of the grand partition function \eqref{zeps} to represent other statistics than Gentile's; as long as the counting operator has eigenvalues 0 or 1. 
A straightforward generalization is to consider situations in which the eigenvalues may be different from 0 or 1. For creation and annihilation operators given in general by \eqref{aadaga}, the eigenvalues are
\begin{equation}
\langle n| \hat{\delta} |n\rangle = \left\{ \begin{array}{ll}
1 & \text{if } n = 0 \\ 
\frac{|\lambda_1 \cdots \lambda_n|^2}{1\cdots n} & \text{if } n \ge 1
\end{array}   \right.
\label{eigend}
\end{equation}
The grand partition function takes the form
\begin{equation}
\mathcal{Z}_{\epsilon} = 1 + \sum_{n=1}^{\infty} \frac{|\lambda_1 \cdots \lambda_n|^2}{n!} x^n.
\label{zeps5}
\end{equation}
Eq.\ \eqref{zeps5} represents a connection between statistics and creation and annihilation operators. It is not difficult to obtain, after a few algebraic steps, the following direct relation between the mean value of $|\lambda_{n+1}|^2$ and the number distribution:
\begin{equation}
\bar{n} = x \overline{|\lambda_{n+1}|^2}.
\end{equation}

Polychronakos \cite{poly} introduced a related approach for exclusion statistics in which the grand partition function for a system of $K$ states with energy $\epsilon$ is written as $\mathcal{Z}(K) = (\mathcal{Z}_\epsilon)^K$, with $\mathcal{Z}_\epsilon = \sum_n P_n x^n$, where $P_n$ are \textit{a priori} probabilities independent of the temperature. In the present context, these probabilities correspond to the eigenvalues of $\hat{\delta}$. In the next sections, the previous relations, mainly Eq.\ \eqref{zeps5}, are applied to different exotic and intermediate statistics. The first case study is the quantum Boltzmann statistics.

\section{Quantum Boltzmann statistics and quons}
\label{squons}

Quons were introduced in order to study possible violations of the Pauli principle \cite{greenberg,mohapatra}; they satisfy the generalized or $q$-commutation relation 
\begin{equation}
a a^\dagger - q a^\dagger a = 1,
\label{qcomm}
\end{equation}
that interpolates between fermions and bosons when $q$ takes values from $-1$ to 1. First we analyze the intermediate case with $q=0$, for which $a a^\dagger = 1$ and
\begin{equation}
\begin{array}{l}
a^\dagger |n\rangle = |n+1\rangle \\ 
a |n\rangle = |n-1\rangle, \quad a|0\rangle=0
\end{array} \quad \text{(quant. Boltzmann)}
\label{qboltz}
\end{equation}
The creation and annihilation operators commute when applied to any non-vacuum number state, this already suggests that the corresponding statistics should be the classical Maxwell-Boltzmann's:
\begin{equation}
\bar{n}_\text{QB} = x,
\end{equation}
that in this context receives the name of Quantum Boltzmann statistics \cite{isakov}. A result that supports the generalization of \eqref{zeps5} beyond Gentile statistics is that it correctly reproduces the Quantum Boltzmann statistics. In this case, we have $\lambda_n=1$ and the grand partition function is
\begin{equation}
\mathcal{Z}_{\epsilon,\text{QB}} = \sum_{n=0}^\infty \frac{x^n}{n!} = \exp\left( x \right).
\end{equation}

Now, let us consider that $q$ takes any value between $-1$ and 1. Using the general expressions for the creation and annihilation operators \eqref{aadaga} combined with the $q$-commutation relation \eqref{qcomm}, the following recursion equation can be obtained: $|\lambda_{n+1}|^2 = 1 + q |\lambda_n|^2$ (see, e.g., \cite{isakov}). Knowing that $\lambda_0 = 0$, we obtain 
\begin{equation}
|\lambda_n|^2 = 1 + q + \cdots + q^{n-1} = \frac{1-q^n}{1-q}.
\end{equation}
This result can be used to obtain the grand partition function from Eq.\ \eqref{zeps5}. It takes the form
\begin{equation}
\mathcal{Z}_{\epsilon,q} = [1-q f_q(x)]^{-1/q},
\label{zfq}
\end{equation}
with
\begin{equation}
f_q(x) = x + (q^3-q)\left[x^3 + (q^3+3q^2+2q-1) x^4 + \cdots \right].
\end{equation}
The number distribution is
\begin{equation}
\bar{n}_q = \frac{f_q'(x)}{x^{-1} - q f_q(x)/x}.
\label{nfq}
\end{equation}
Since for $q=0, -1$ or 1 we have that $f_q(x)=x$, we recover the Quantum Boltzmann, Fermi-Dirac and Bose-Einstein statistics for that cases.

Isakov proposed an ansatz for the evaluation of the number distribution for quons  (see Eq.\ (67) in Ref.\ \cite{isakov}), and obtained
$$
\frac{1}{x^{-1} - q},
$$
that also recovers the cases of fermions, bosons and classical particles for $q=-1, 1$ and 0. But this result has the following drawback. Let us consider $q=-1/p$, where $p$ is a positive integer. In the limit of small energy, the number distribution tends to a maximum possible value equal to $p$. This is correct for $p=1$; for larger values of $p$ this limit means that the creation operator applied to $|p\rangle$ should be zero, but this is not actually the case, since $a^\dagger |p\rangle = \lambda^*_{p+1}|p+1\rangle \neq 0$ for $p \geq 2$. This simple number distribution actually corresponds to the Polychronakos statistics, that is analyzed in the next section.

The commutation relation \eqref{qcomm} is actually a particular case of two-parameter quantum algebras \cite{chakra,tuszynski}. I consider one more example of these $q$ (or $qp$) deformed algebras. The commutation relation for $q$-bosons is
\begin{equation}
a a^\dagger - q a^\dagger a = q^{-\hat{n}},
\label{qcomm2}
\end{equation}
with $q>0$, for which we have 
\begin{equation}
|\lambda_n|^2 = \frac{q^n - q^{-n}}{q - q^{-1}}.
\end{equation}
Now, the coefficients needed to evaluate the grand partition function in \eqref{zeps5}, $|\lambda_1 \cdots \lambda_n|^2/n!$, diverge for increasing values of $n$ faster than $e^n$. We can not obtain a convergent series for the grand partition function in this case, unless for $q=1$, corresponding to bosons. Starting from \cite{tuszynski}, several papers have analyzed statistics and thermodynamic properties of $q$-deformed algebras, including this last case. According to \cite{dai}, those results are incorrect since the unjustified approximation $\overline{n^2} \simeq \bar{n}^2$ (or $\overline{q^n} \simeq q^{\overline{n}}$) is generally used.

\section{Polychronakos statistics}
\label{spoly}

Based on the fractional exclusion statistics introduced by Haldane \cite{haldane}, Polychronakos \cite{poly} proposed an alternative definition that has the following advantages. When the grand partition function for a single level is written as $\mathcal{Z}_\epsilon = \sum_n P_n x^n$, any $P_n$ takes positive values independent of the number of states $K$, a maximum occupancy number results for fermionic cases, and the expressions for thermodynamic quantities turn out to be analytic. Polychronakos statistics is based on a variation of the exclusion statistics for a system of $K$ states of energy $\epsilon$: the inclusion of the first particle leaves $K- \alpha$ states for the second, the inclusion of the second leaves $K-2\alpha$, and so on. The combinatorial formula for putting $n$ particles in $K$ states is
\begin{equation}
W = \frac{K(K-\alpha)\cdots (K - (n-1)\alpha)}{n!}.
\end{equation}
The corresponding number distribution is \cite{poly}:
\begin{equation}
\bar{n}_\text{P} = \frac{1}{x^{-1} + \alpha}.
\end{equation}
The grand partition function is
\begin{equation}
\mathcal{Z}_{\epsilon,\text{P}} = (1+\alpha x)^{1/\alpha},
\end{equation}
and a series expansion gives
\begin{equation}
P_n = \frac{1}{n!}\prod_{m=0}^{n-1} (1 - m\alpha)
\end{equation}
for $n \geq 1$, and $P_0=1$. For $\alpha=0$ we have the Boltzmann distribution. For $\alpha < 0$ we have the so-called bosonic sector, and all probabilities are positive. For positive values of $\alpha$, only $\alpha=1/p$, with $p$ a positive integer, are considered, so that the probabilities are positive up to $n=p$ and vanish for $n\geq p+1$.

The connection with creation and annihilation operators becomes immediate when comparing with Eq.\ \eqref{zeps5}, since the coefficients in \eqref{zeps5} are equal to $P_n$. Assuming that $\lambda_n$ are real, we obtain $\lambda_n = \sqrt{1 - (n-1)\alpha}$, and the creation and annihilation operators behave as
\begin{align}
a^\dagger |n\rangle &= \sqrt{1-n\alpha}\, |n+1\rangle \nonumber \\ 
a |n\rangle &= \sqrt{1-(n-1)\alpha}\, |n-1\rangle,
\end{align}
including the vacuum condition $a |0\rangle = 0$. The previous relations correspond to bosons for $\alpha=-1$, classical particles for $\alpha=0$, and fermions for $\alpha=1$.

\begin{figure*}
	\includegraphics[width=.47\textwidth]{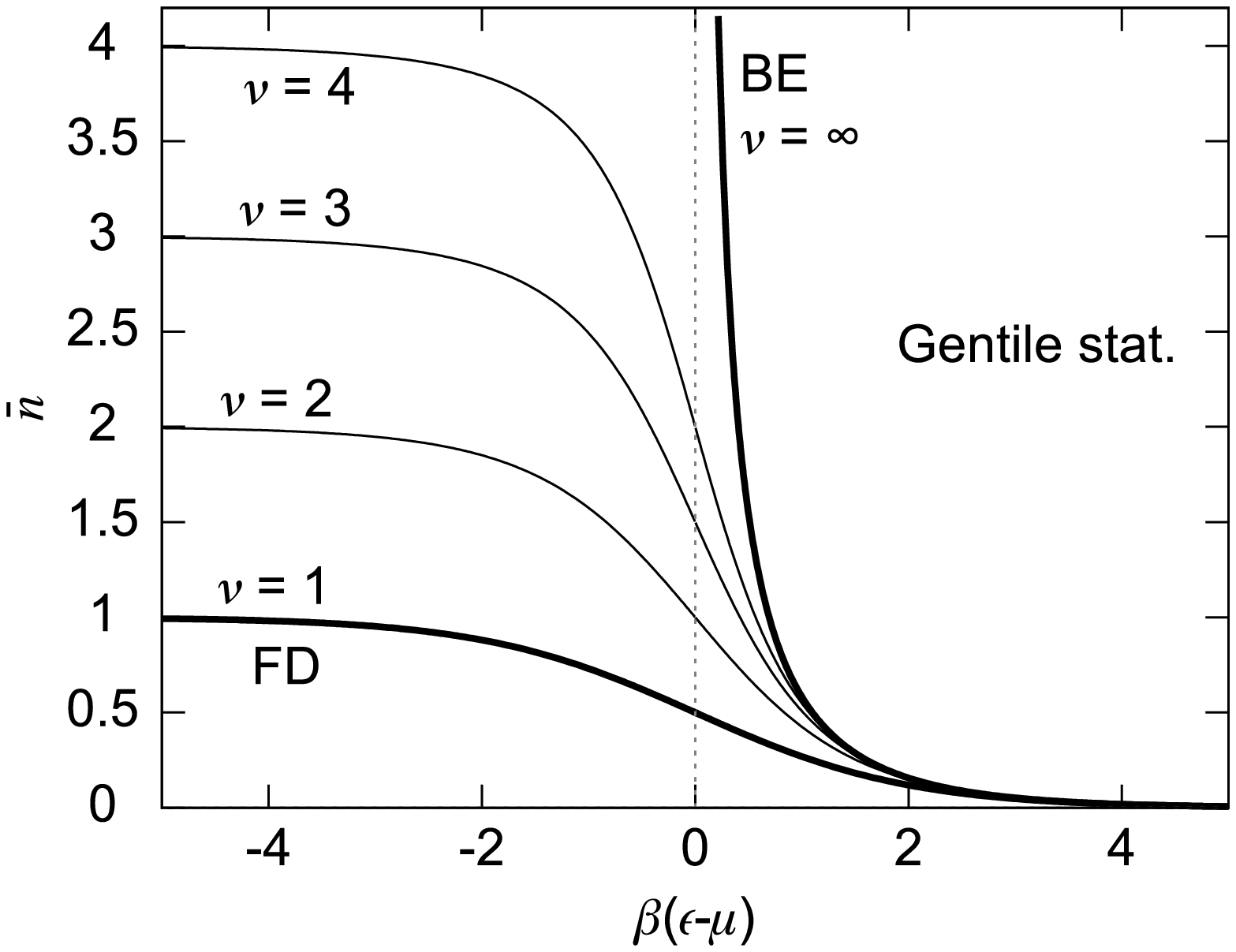}\hfill
	\includegraphics[width=.47\textwidth]{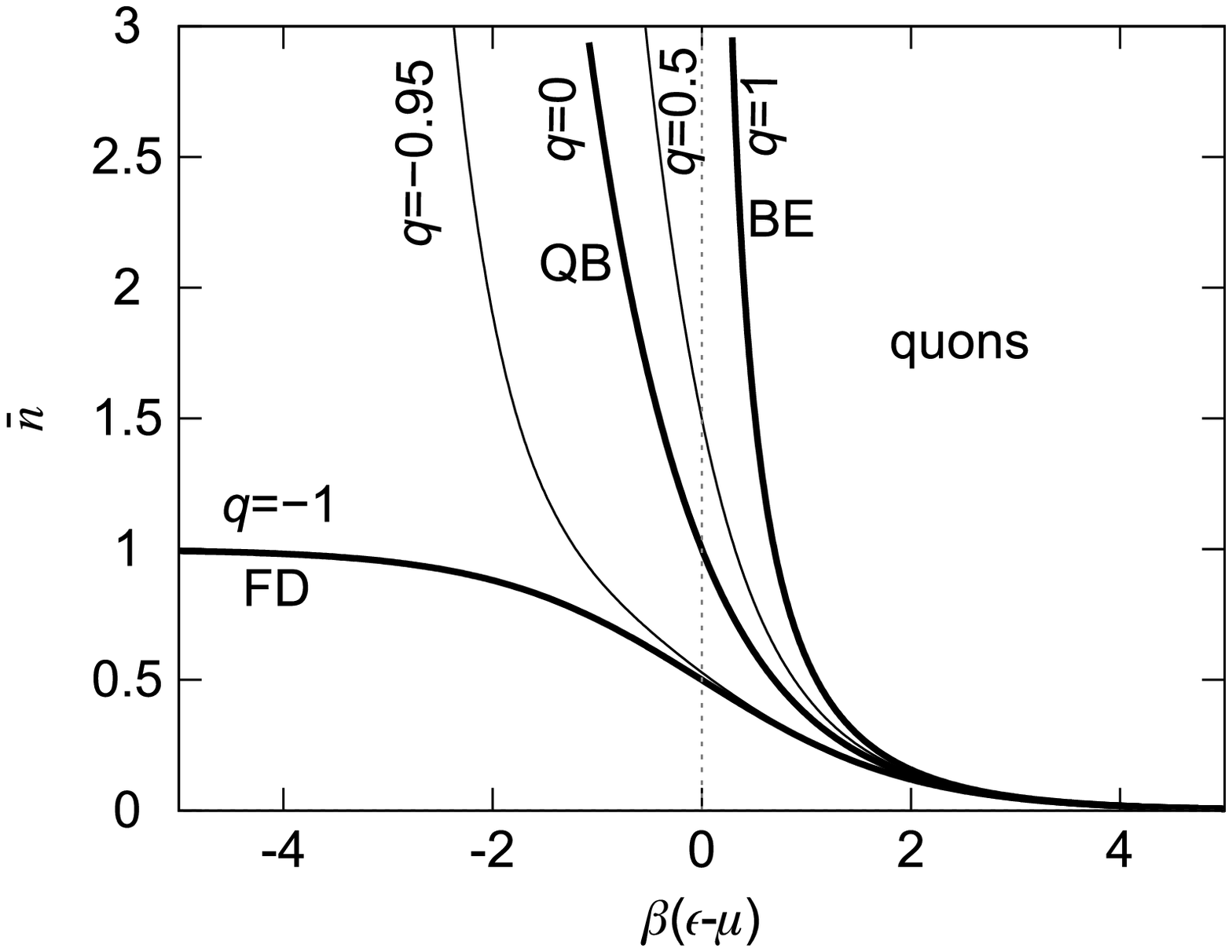}
	\includegraphics[width=.47\textwidth]{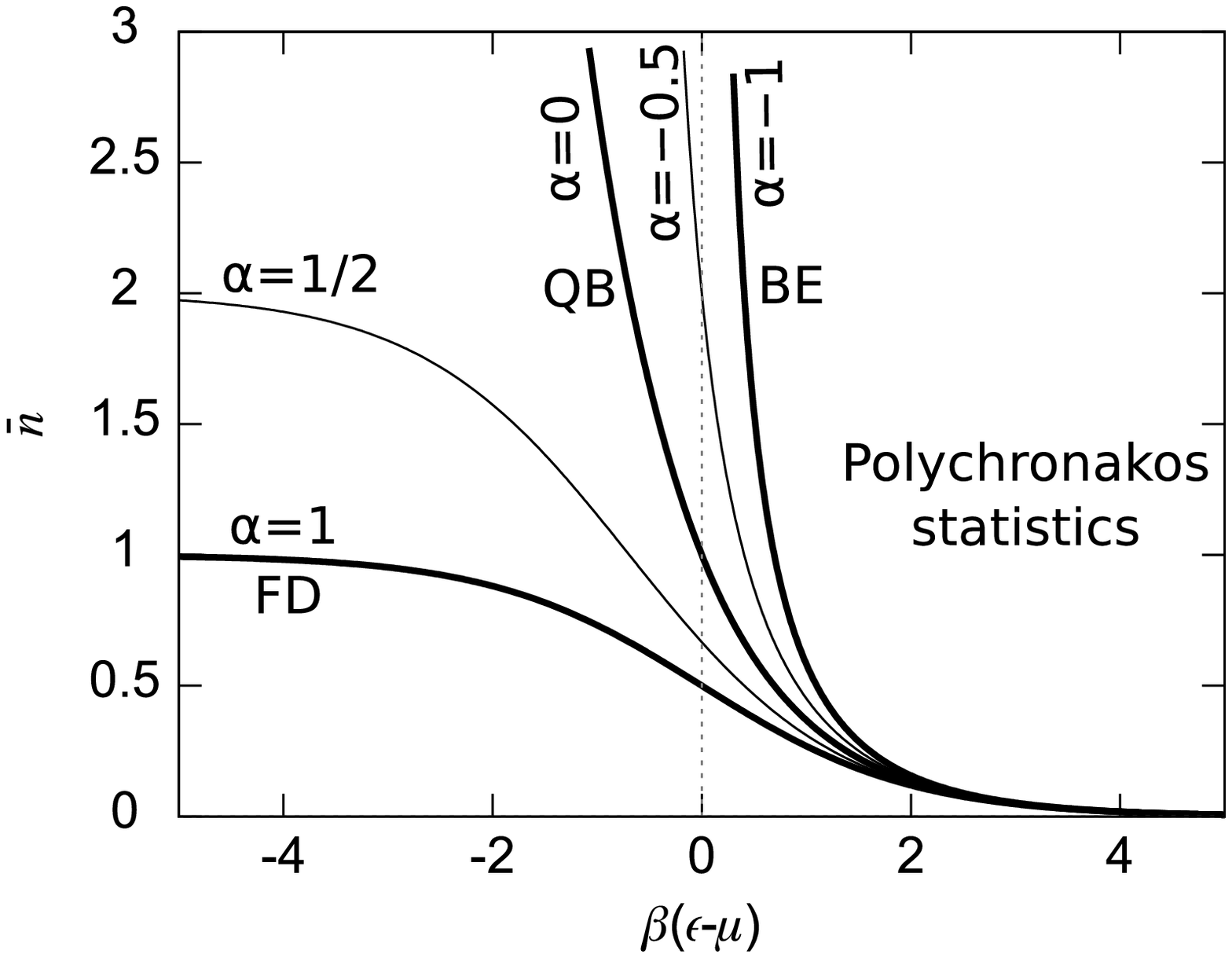}\hfill
	\includegraphics[width=.47\textwidth]{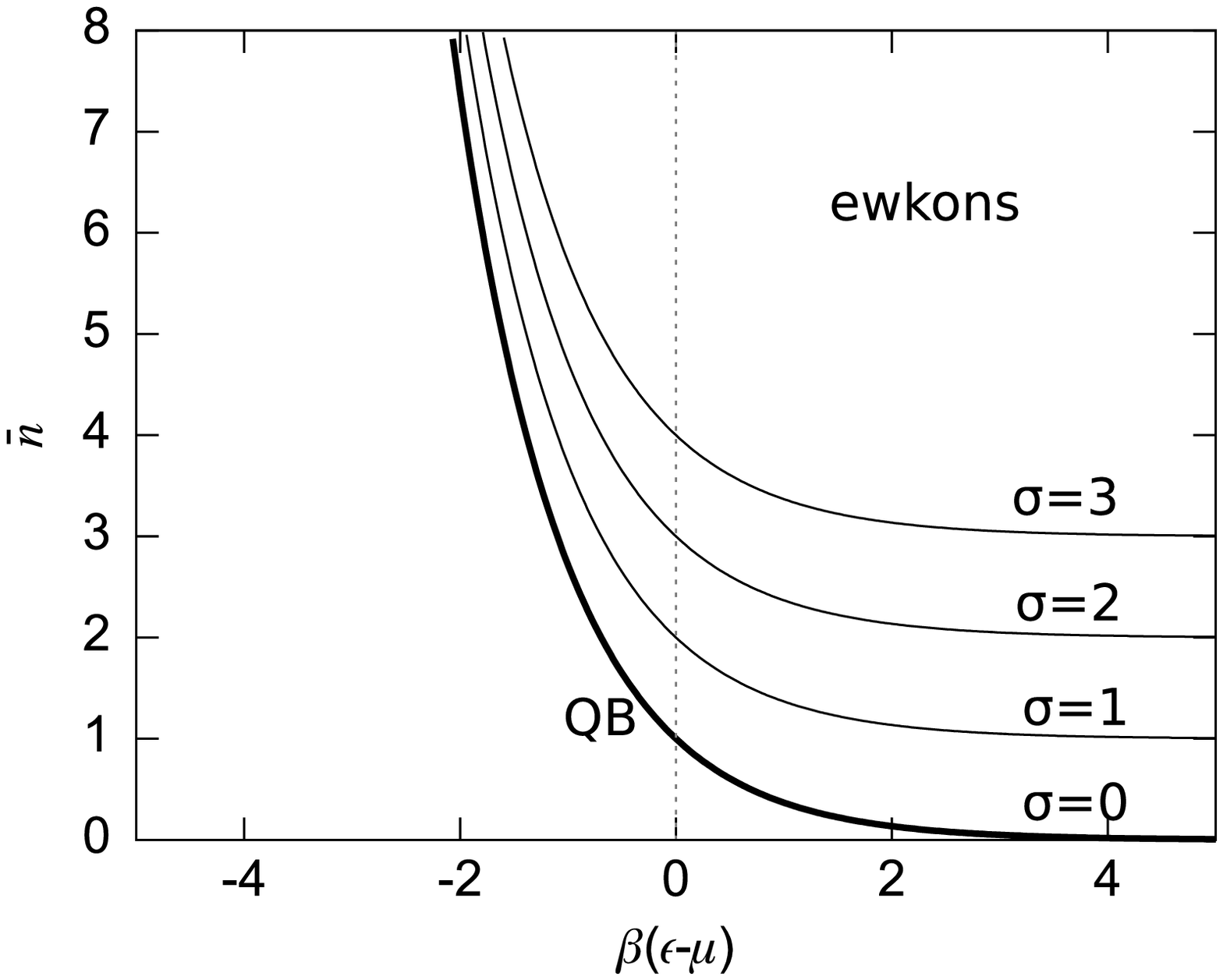}
\caption{Number distribution $\bar{n}$ against $\beta(\epsilon-\mu)$ for different statistics are shown side by side in order to facilitate comparison. From left to right and from top to bottom: Gentile statistics, quons, Polychronakos statistics and ewkons. Fermi-Dirac (FD), Bose-Einstein (BE) and Quantum Boltzmann (QB) statistics are highlighted with thicker stroke. \label{figstat}}	
\end{figure*}

\section{Ewkons}
\label{sewkons}

A nonlinear Fokker-Planck equation for the diffusion of noninteracting particles in energy space was proposed recently in Ref.\ \cite{hoyusist}. The equation was based on previous work on classical particles with effective potentials that reproduce quantum statistics \cite{kania,kania2,toscani,suarez2}. Assuming that noninteracting particles have free diffusion, then classical, Bose-Einstein, and Fermi-Dirac statistics are derived. And also an additional statistics for particles called ewkons. The ewkon statistics is given by
\begin{equation}
\bar{n}_\text{ewk} = \sigma + x,
\label{ewkonstat}
\end{equation} 
It is equal to the Boltzmann distribution displaced a fixed quantity $\sigma$, with $\sigma$ a positive integer.  Ewkons have a non vacuum ground state. The problem of divergent thermodynamic quantities is addressed in Subsect.\ \ref{aplew}.

Before analyzing ewkons, let us consider a further generalization of the reasonings of Sect.\ \ref{gent}, where the connection between statistics and creation and annihilation operators is presented in Eq.\ \eqref{zeps5}. Let us go back to the definition \eqref{zeps2} of the grand partition function. We obtain Gentile statistics by a restriction of the sum to the allowed values of $n$. Now let us take into account that the restriction is not only on the maximum value of $n$ but also on the minimum. I consider a non vacuum ground state with $\sigma$ particles: $|\sigma \rangle$. Then, considering that $\mathcal{Z}_\epsilon = \sum_{n=\sigma}^{\nu} x^n$, with $\nu > \sigma$, the number distribution is
\begin{equation}
\bar{n} = \frac{1}{x^{-1}-1} - \frac{\nu+1-\sigma}{x^{-(\nu+1-\sigma)}-1} + \sigma.
\end{equation}
For example, for $\nu\rightarrow \infty$, it reproduces the Bose-Einstein distribution displaced a quantity $\sigma$. 

The vacuum condition has to be replaced by $a |\sigma \rangle = 0$, or $\lambda_\sigma=0$. The counting operator for this shifted Gentile statistics has eigenstates $|n\rangle$ with eigenvalue 1 if $\sigma \le n \le \nu$ and 0 if $n \ge \nu+1$. Reproducing the steps of Sect.\ \ref{gent}, expression \eqref{eigend} is modified in the following way:
\begin{equation}
\langle n| \hat{\delta} |n\rangle = \left\{ \begin{array}{ll}
0 & \text{if } n \le \sigma -1 \\
1 & \text{if } n = \sigma \\ 
\frac{|\lambda_{\sigma+1} \cdots \lambda_n|^2}{(\sigma+1)\cdots n} & \text{if } n \ge \sigma+1
\end{array}   \right.,
\label{eigend2}
\end{equation}
This definition reduces to \eqref{eigend} for $\sigma=0$. Let us note that $\sigma$ must be non-negative in order to avoid the indeterminacy $\lambda_0/0$ (this precludes the possibility of genkons, mentioned in Ref.\ \cite{hoyusist}).  The grand partition function takes the form
\begin{equation}
\mathcal{Z}_{\epsilon} = x^{\sigma} + \sum_{n=\sigma+1}^{\infty} \frac{|\lambda_{\sigma+1} \cdots \lambda_n|^2}{(\sigma + 1)\cdots n} x^n.
\label{zeps6}
\end{equation}
Now we can analyze ewkons. The corresponding grand partition function is
\begin{equation}
\mathcal{Z}_{\epsilon,\text{ewk}} = e^x\,x^\sigma.
\end{equation}
Assuming that $\lambda_n$ is real, a series expansion gives $\lambda_n = \sqrt{n}/\sqrt{n-\sigma}$, and the creation and annihilation operators for ewkons are
\begin{equation}
\begin{array}{l}
a^\dagger |n\rangle = \sqrt{\frac{n+1}{n+1-\sigma}}\,|n+1\rangle \\ 
a |n\rangle = \sqrt{\frac{n}{n-\sigma}}\,|n-1\rangle, \quad
a|\sigma\rangle = 0
\end{array} \quad \text{(ewkons)}
\end{equation}
If $\sigma=0$, we recover the operators for the Quantum Boltzmann distribution \eqref{qboltz}.

The observed accelerated expansion of the universe is accounted by the dark energy, that should have a negative relation between pressure and energy density \cite{hogan,Planck2015,Caldwell}; also, it is homogeneously distributed in whole space.  Ewkons have this two properties that make them suitable to describe dark energy: since any energy level should have at least $\sigma$ particles, they are spread in whole space (assuming a homogeneous number of states per unit volume); and, on the other hand, they have a negative relation between pressure and energy density.

\subsection{Application to dark energy}
\label{aplew}

In order to obtain thermodynamic properties of an ideal gas of ewkons of mass $m$, I assume that the energy gaps are small enough to consider a continuous energy spectrum and introduce a density of states $g(\epsilon)$. Following the same procedure used for fermions and bosons, states are determined by a wavevector $\textbf{k}$ in a volume $V$; in a nonrelativistic gas, they have energy $\epsilon = \hbar^2 k^2/2m + mc^2$ and the density of states is $g(\epsilon') = g_d V 2\pi (2m)^{3/2} \epsilon'^{1/2}/h^3$, where $\epsilon'= \epsilon-mc^2$ and $g_d$ is a degeneracy factor; see, e.g., Ref. \cite[p. 33]{pathria}. In order to avoid divergences in the total energy or number of particles, it is necessary to introduce an ultraviolet cutoff. There is a maximum value for the energy, $\epsilon_\text{max}$, such that $g(\epsilon')=0$ for $\epsilon' > \epsilon_\text{max}$. I assume that $y=k_B T/ \epsilon_\text{max} \ll 1$. Using the grand partition function, the expressions for the energy density and pressure are:
\begin{align*}
\rho &= \frac{1}{V} \int d\epsilon\; g(\epsilon)\, \epsilon\, \bar{n}_\text{ewk} \\
P &= \frac{1}{V\beta} \int d\epsilon\; g(\epsilon) \ln \mathcal{Z}_{\epsilon,\text{ewk}}.
\end{align*}
The results are:
\begin{align}
\rho &= \epsilon_\text{max}^{5/2}  \frac{\sigma g_d \sqrt{2}\, m^{3/2}}{5\pi^2 \hbar^3}\left[1 + \frac{5}{3}\frac{m c^2}{\epsilon_\text{max}} + \mathcal{O}(y^{5/2})\right],
\label{endensity} \\
P &= \epsilon_\text{max}^{5/2} \frac{\sigma g_d \sqrt{2}\, m^{3/2}}{5\pi^2 \hbar^3}\left[-1 + \frac{5}{3}\frac{\mu-mc^2}{\epsilon_\text{max}} + \mathcal{O}(y^{5/2})\right].
\end{align}
Le us note that, independently of the temperature, the rest energy term in the energy density, $\frac{5}{3}m c^2/\epsilon_\text{max}$, is not dominant (as is the case for fermions) since all energy levels (up to $\epsilon_\text{max}$) should be occupied. 
The relation between pressure and energy density, $w_\text{ewk}=P/\rho$, i.e. the cosmological equation of state for ewkons, is 
\begin{equation}
w_\text{ewk} = -1 + \frac{5}{3} \frac{\mu}{\epsilon_\text{max}} + \mathcal{O}(y^{5/2}),  
\label{wewk} 
\end{equation}
where it was assumed that $\frac{\mu-mc^2}{\epsilon_\text{max}} \ll 1$. Then, $w_\text{ewk}$ is equal to $-1$ plus a quantity of order $\mu/\epsilon_\text{max}$. The main current models for dark energy, cosmological constant and quintessence, include a negative pressure. This result is in agreement with recent observations of the present value of $w$, mainly dominated by dark energy, that establish an upper bound $w < -0.94$ at 95\% confidence level; see Table 3 in Ref.\ \cite{Planck2015}. Also, Eq.\ \eqref{endensity} is in accordance with the observation that dark energy density remains almost constant as the universe expands (see, e.g., \cite{wang}), assuming that $\epsilon_\text{max}$ is independent of the universe scale.


Using that the dark energy density is approximately equal to $4\; 10^9$ eV/m$^3$ \cite{kowal}, we can obtain
\begin{equation}
m \lesssim 0.006\ \text{eV}/c^2 
\end{equation}
assuming $\sigma \ge 1$, $g_d \ge 1$ and $\epsilon_\text{max}/ mc^2>1$. This small upper bound for the mass connects the present approach with quintessence theories, where a nearly massless scalar field accounts for the dark energy density \cite{caldwell} (values of the mass between $0.00243$ and $0.00465$ eV$/c^2$ are used in \cite{caldwell}).

\section{Conclusions}
\label{conclusions}

Statistics of systems composed by non-interacting particles is obtained form the single level grand partition function $\mathcal{Z}_\epsilon = \mathrm{tr}\, e^{-\beta\,(\epsilon-\mu) \hat{n}}$. The definition of $\mathcal{Z}_\epsilon$ implies that only Gentile statistics is possible \cite{dai}. It is interesting, however, to extend this definition in order to include other statistics, most noticeable the Boltzmann or Quantum Boltzmann statistics for $\lambda_n=1$. Such an extension would be relevant for several exotic statistics present in the literature that, for example, transfer the effects of interactions to features of creation and annihilation operators \cite{anghel}. I introduced a counting operator $\hat{\delta}$ that has eigenvalue 1 for the number states that are allowed, and 0 otherwise, so that its inclusion in the definition of $\mathcal{Z}_\epsilon$ does not have any effect on Gentile statistics. The counting operator can be written in terms of creation and annihilation operators. It is a natural extension to consider situations in which the eigenvalues of $\hat{\delta}$ are not only 0 or 1. This extension results in a number distribution that is consistent with expected features for some properties of creation and annihilation operators. For example, it reproduces the Quantum Boltzmann statistics for $\lambda_n=1$. The number distribution has an upper bound equal to $p$ if $a^\dagger |p\rangle = 0$ and is not necessarily bounded if there is no $p$ such that the previous condition holds. 

Besides Gentile and Quantum Boltzmann statistics, the procedure was applied to quons, $q$-bosons, Polychronakos statistics and ewkons; see Fig.\ \ref{figstat}. For quons, the number distribution was obtained from the commutation relation; for $q$-bosons, the grand partition function turns out to be divergent. For Polychronakos statistics and ewkons, the creation and annihilation operators were obtained from the grand partition function. In the case of ewkons, the statistics was deduced in Ref.\ \cite{hoyusist} from the condition of free diffusion in energy space; condition that is also fulfilled by fermions, bosons and classical particles. An ideal gas of ewkons has negative pressure and a cosmological equation of state similar to $-1$ plus a term of order $\mu/\epsilon_\text{max}$, see Eq.\ \eqref{wewk}; these features make them appropriate for the description of dark energy. They also provide a qualitative understanding of  the fact that accelerated expansion was not always present, but started when the universe became less dense, situation that corresponds to an ewkon's chemical potential small enough to satisfy the condition imposed by the Friedmann equation to have accelerated expansion: $w < -1/3$.

\begin{acknowledgments}
I acknowledge useful discussions with Pablo Sisterna and Héctor Mártin. This work was partially supported by Consejo Nacional de Investigaciones Científicas y Técnicas (CONICET, Argentina, PIP 0021 2015-2017).
\end{acknowledgments}

\end{document}